\documentclass[12pt]{amsart}
\usepackage{amsaddr}

\usepackage{booktabs, cite, graphicx, multirow}
\usepackage{amsthm, amsmath, amsfonts, amssymb, array}
\usepackage[linesnumbered, algoruled, boxed, lined]{algorithm2e}

\usepackage{geometry}
\usepackage{caption}
\newcommand{\rr}{\mathbf{r}}
\newcommand{\xx}{\mathbf{x}}
\newcommand{\nn}{\mathbf{n}}
\newcommand{\uu}{\mathbf{u}}

\newcommand{\ES}{\mathcal{S}^2}
\newcommand{\nmax}{{n_{\rm max}}}
\newcommand{\cin}{{c_{n,l}^{in}}}
\newcommand{\cout}{{c_{n,l}^{out}}}

\newcommand{\NBR}{\mathcal{N}_{\bf r}}

\graphicspath{{./images/}}

\title{Towards a quantitative assessment of neurodegeneration in Alzheimer's disease}
\author{Oleg Michailovich and Rinat Mukhometzianov}
\address{The Department of Electrical and Computer Engineering, University of Waterloo, Waterloo, ON N2L 3GL, Canada}
\date{}

\begin{document}

\maketitle

Submitted to \textbf{IEEE Access}

\section*{Abstract}
Alzheimer's disease (AD) is an irreversible neurodegenerative disorder that progressively destroys memory and other cognitive domains of the brain. While effective therapeutic management of AD is still in development, it seems reasonable to expect their prospective outcomes to depend on the severity of baseline pathology. For this reason, substantial research efforts have been invested in the development of effective means of non-invasive diagnosis of AD at its earliest possible stages. In pursuit of the same objective, the present paper addresses the problem of the quantitative diagnosis of AD by means of Diffusion Magnetic Resonance Imaging (dMRI). In particular, the paper introduces the notion of a {\it pathology specific imaging contrast} (PSIC), which, in addition to supplying a valuable diagnostic score, can serve as a means of visual representation of the spatial extent of neurodegeneration. The values of PSIC are computed by a dedicated deep neural network (DNN), which has been specially adapted to the processing of dMRI signals. Once available, such values can be used for several important purposes, including stratification of study subjects. In particular, experiments confirm the DNN-based classification can outperform a wide range of alternative approaches in application to the basic problem of stratification of cognitively normal (CN) and AD subjects. Notwithstanding its preliminary nature, this result suggests a strong rationale for further extension and improvement of the explorative methodology described in this paper.

\smallskip
\noindent \textbf{Keywords:} diffusion MRI, deep learning, convolutional neural networks, early diagnosis, Alzheimer's disease. 

% ---------------------------------------------------------------- SECTION 1 ------------------------------------------------------------------
\section{Introduction}\label{Sec1}
The world population is steadily ageing, and with advanced age comes a higher risk of dementia. At the present time, the dementia of Alzheimer's type, or {\it Alzheimer's disease} (AD), accounts for almost two-thirds of all prevalent cases of dementia in the elderly. AD is an irreversible, progressive disease that slowly destroys memory and other cognitive domains, eventually leaving the patient bedridden. The course of AD pathology is likely to span around two to three decades \cite{Kumar:2015aa}. Unfortunately, by the time when the first symptoms emerge, it is usually too late to save the brain. For this reason, over the last two decades, considerable efforts have been directed towards finding effective means of the earliest possible diagnosis of AD \cite{AD2018}.

The current arsenal of methods for quantitative diagnosis of AD is impressively broad, ranging from advanced proteomics to state-of-the-art neuroimaging. In the latter case, particularly promising results have been demonstrated by both nuclear and Magnetic Resonance Imaging (MRI) \cite{Frisoni:2010aa, Weiner:2012aa, AD2018}.

Among various methods of MRI, {\it diffusion} MRI (dMRI) is exceptional for its unique ability to generate imaging contrast based on the microscopic (rather than macroscopic) properties of neurological tissue, which makes it singularly fit for the task of detecting the earliest signs of neurodegeneration \cite{Basser:1994aa, Bihan:1986aa}. This ability of dMRI has been investigated in a number of studies \cite{Acosta:2012aa, Amlien:2014aa, Schouten:2017aa, Zhang:2014aa, Mayo:2018aa}, which predominantly focused on the problem of classification (aka {\it stratification}) of three groups of subjects, {\it viz.} cognitively normal (CN) subjects, AD subjects, and the subjects diagnosed with {\it mild cognitive impairment} (MCI). Note that the latter is broadly recognized as a prodromal condition that frequently heralds the onset of ``full-blown" AD \cite{Petersen:2016aa, Edmonds:2019aa}. 

In virtually all earlier studies on dMRI-based stratification of AD, MCI, and CN subjects, the protocol of choice has been {\it Diffusion Tensor Imaging} (DTI) \cite{Bihan:1986aa, Basser:1994aa, Mori:2014aa}. The latter is known to provide an adequate characterization of diffusion dynamics in the white matter associated with {\it non-crossing} bundles of neural fibre tracts. Unfortunately, its dependence on Gaussian modelling curbs the ability of DTI to delineate more complex diffusion processes, e.g., within {\it crossing} fibres \cite{Assaf:2008aa, Jones:2010aa}. It is thus no wonder that, even though many DTI metrics have demonstrated considerable sensitivity across multiple brain regions, the most consistent findings have been confined to the corpus callosum \cite{Bozzali:2002aa, Acosta:2012aa, Lee:2013aa, Amlien:2014aa, Teipel:2014aa, Hawkins:2015aa, Mayo:2017aa, Schouten:2017aa, Mayo:2018aa}.  At the same time, few DTI studies have been able to stratify CN, AD, and MCI subjects based on DTI analysis of the medial-temporal white matter, which is known to be abundant in both crossing and ``kissing" fibre tracts. The problem here has obviously been in the intrinsic modelling limitations of DTI, which is rather discouraging out-turn in view of the known involvement of the above region in the early stages of neuropathological AD \cite{Englund:1988aa}. 

The limitations of DTI have prompted the development of more advanced methods of dMRI, among which Neurite Orientation Dispersion and Density Imaging (NODDI) is considered to be one of the most comprehensive approaches to quantitative characterization of cerebral diffusion \cite{Zhang:2012aa}. Naturally, several studies have investigated the applicability of NODDI to early diagnosis of AD. However, when trying to correlate the spatial distribution of NODDI metrics with histopathological evidence of AD, it was observed in \cite{Colgan:2016aa} that NODDI offered somewhat marginal advantages over DTI. A similar conclusion was reached in \cite{Slattery:2015aa}, in which NODDI was used in application to diagnosis of young-onset AD.

Needless to say, DTI and NODDI are only two specific examples among a wide range of methods available under the umbrella of dMRI. However, regardless of their specific modelling assumptions, all these methods share a tendency to produce more accurate results at the expense of higher complexity of parametrization. At the same time, the use of parametric spaces of progressively higher dimensionality requires a proportional increase in the number of data points, which might not always be possible due to practical constraints. Thus, given the melange of available protocols and models, the question of {\it which of the existing dMRI methods is ``the best" for early diagnosis of AD} appears to be rather non-trivial.

Before going any further, it is important to note that not all methods of dMRI are equally feasible from the viewpoint of clinical implementation. In particular, for practical reasons, the typical duration of a clinical dMRI examination rarely exceeds 15-20 mins. This constraint puts a strict upper bound on the amount of acquirable data, and, consequently, on the maximal order of numerically stable parametrization. For this reason, most of the studies on the dMRI-based diagnosis of AD have predominantly relied on DTI. Despite its numerous limitations, DTI remains ``the method of choice" in many ongoing studies thanks to the minimality of its technical requirements and its time efficiency.

It is also worthwhile noting that, in quotidian exchanges, the term ``DTI" is usually used in two different connections. In particular, it could refer to the Gaussian (i.e., 2$^{\rm nd}$-order tensor) diffusion model which lies in the foundation of DTI analysis. This model is described by a total of seven parameters\footnote{The DTI model is parametrized by a symmetric $3\times 3$ diffusion tensor having six independent entries. An additional parameter is required to account for the effect of normalization.}, which could be estimated based on a minimum of seven independent measurements. Although the practical number of diffusion measurements normally exceeds this low bound, their acquisition is still rare to require more than 15 mins of scanning time, which makes such imaging protocols {\it clinically feasible}. It is probably due to the association between the low dimensionality of DTI modelling and its dependence on a relatively small number of measurements that the term ``DTI" has also been used to refer to diffusion data comprised of a comparatively small number of diffusion encodings. Hence, to avoid misconception, the terms DTI {\it data} and DTI {\it modelling} need to be distinguished.

For the reasons explained above, virtually all clinical dMRI data could be characterized as ``DTI". From the practical point of view, therefore, it seems reasonable to restrict the scope of available dMRI models to only those which could be reliably fitted based on clinical DTI data. Unfortunately, this would have mainly left us with low-parametric models of a descriptive power similar to that of DTI. This out-turn reveals a critical methodological predicament, where the use of more advanced models is fraught with estimation artefacts, whereas suppressing such artefacts by including additional measurements is precluded by practical constraints. In such conditions, the use of low-parametric dMRI models becomes the only option, which, unfortunately, comes at the cost of reduced accuracy.

A particularly promising way to improve over the performance of low-parametric dMRI modelling is offered by {\it data-driven inference} and, in particular, by its recent realization in the theory of {\it Deep Learning} (DL) \cite{LeCun:2015aa, Goodfellow:2016aa}. The modern methods of DL make it possible to discern subtle and complex dependencies in experimental data, which would have been impossible to describe in mechanistic terms. In this case, the actual (unknown) model is replaced by its phenomenological representation in terms of a {\it Deep Neural Network} (DNN) that is capable of ``learning" to {\it predict} future outcomes based on past observations.

Although the idea of using DL in the imaging-based diagnosis of AD is not original, much of the work along this direction has mainly focused on {\it structural} MRI. The latter has proven instrumental in assessing the cerebral atrophy due to AD in both cortical and subcortical regions of the brain \cite{Bhagwat:2019aa, Jain:2019aa, Wang:2019bb, Choi:2018aa, Lian:2018aa}. In longitudinal studies, the substantial potential for prediction of ensuing cognitive decline in MCI and AD subjects has been demonstrated through the use of recurrent DNNs \cite{Ghazi:2019aa, Lee:2019aa, Wang:2019aa, Aghili:2018aa, Lin:2018aa, Kawahara:2017aa}, both with and without augmenting the structural MRI data with other sources of diagnostic information \cite{Lee:2019aa, Zhou:2019aa, Forouzannezhad:2018aa, Lu:2018aa}. However, the application of DL to a dMRI-based diagnosis of AD remains a barely tapped in area of research, notwithstanding the abundant evidence of its successful use in other applications \cite{Golkov:2016aa, Koppers:2016aa}. Accordingly, the primary goal of this work has been to leverage the combined power of dMRI and DL towards the early diagnosis of AD. 

The key idea of the proposed methodology is built around the notion of {\it pathology specific imaging contrast} (PSIC). Similarly to other imaging-based markers, PSIC is a scalar score that indicates the presence of a suspected pathology (such as, e.g., AD). However, instead of characterising an entire dataset, the values of PSIC are computed at each spatial coordinate within a specified region-of-interest (ROI). In this way, PSIC can serve as a local indicator of the degree to which AD pathology affects various anatomical sites.

Once available, the values of PSIC can be converted into regional statistics, which can, in turn, be used for the purpose of subject classification. In the case of multiple ROIs, the performance of such classification could be further improved through exploring the interdependencies between different regional statistics, which are known to undergo sizeable changes in AD \cite{Maryam:2017aa}. Furthermore, the local PSIC values can be displayed in superposition with anatomical images in the form of a {\it contrast}. In this way, PSIC also offers a means for visual analysis of the extent and severity of suspected pathology.
  
In this work, the values of PSIC have been generated by a dedicated DNN. Although the overall architecture of the DNN is based on a standard feed-forward configuration, its principal operations (such as, e.g., convolution, pooling, etc.) have been properly adjusted to the physical and analytical properties of diffusion signals. The adjustment made it possible to minimize the number of network parameters and, consequently, to reduce the amount of training data substantially. In particular, the results of this paper have been obtained based on only 40 dMRI datasets.

It is important to emphasize that the results reported here should be regarded as neither exhaustive nor final, but rather describing a novel concept which admits many possible extensions and improvements. For this reason, no attempts have been made to ``push the limits" of the proposed method by testing its performance in deliberately difficult scenarios. Instead, the present experimental study has been deliberately limited to the basic task of stratification of CN and AD subjects, while focusing instead on a comparative performance between the proposed method and a range of existing alternatives.

The remainder of this paper is organized as follows. The principal idea of the proposed method is described in Section 2, while Section 3 introduces some necessary technical preliminaries, followed by a description of the proposed network design in Section 4. Subsequently, Section 5 provides details on the experimental setup, study data, and performance metrics used in this work, while experimental results are summarized in Section 6. Finally, Section 7 concludes the paper with a discussion of its main findings along with an outline of possible directions for future research.

% ---------------------------------------------------------------- SECTION 2 ------------------------------------------------------------------
\section{Principal idea and data structure}\label{Sec2}
The experimental study of this work has been based on the dMRI data available through the continuing efforts of the Alzheimer's Disease Neuroimaging Initiative (ADNI)\footnote{For more details on various ADNI programs, visit {\tt adni.loni.usc.edu}.}. Over the last few years, the ADNI database has been extended to include dMRI data acquired by means of relatively advanced protocols. For the purposes of this paper, however, we used the dMRI data collected during an earlier phase of the ADNI study -- known as ADNI-II. At that time, the data acquisition relied on more standard protocols which are more common in present-day clinical DTI. Thus, working with the earlier data should provide a more objective demonstration of the practical value of the proposed methodology. 

In what follows, we consider a typical setting in which a DTI dataset consists of $K$ diffusion-encoded MRI volumes which encode the values of apparent diffusivity along different spatial orientations. Such data are usually acquired at a fixed level of diffusion sensitization controlled by the $b$-value \cite{Mori:2014aa}\footnote{For relatively large $K$, this acquisition scheme is also known as {\it High Angular Resolution Diffusion Imaging} (HARDI) \cite{Tuch:2004}.}. In particular, in the case of ADNI-II data, the number of diffusion encodings was set to $K=41$, with $b = 1000$ s/mm$^2$.   

Formally, DTI signals can be considered to be functions of both spatial and spherical coordinates. In practice, the spatial coordinate is sampled over a regular Cartesian grid $\Omega : = \{ \nn =(n_1, n_2, n_3) \mid 0 \le n_i < N_i, \, i =1, 2, 3\}$ which represents the (anatomical) image domain. On the other hand, the process of diffusion encoding restricts the signal values to $K$ points $\{\uu_k \mid \| \uu_k \| = 1\}_{k=1}^K$ over the unit sphere $\ES$ in the diffusion $\bf q$-space. Thus, from a practical point of view, a DTI dataset can be viewed as a 4-D numerical array of size $N_1 \times N_2 \times N_3 \times K$, with three ``anatomical" and one ``diffusion" dimension.

For some $\rr \in \Omega$, let $\mathcal{N}_\rr$ be a symmetric neighbourhood of $\rr$ consisting of all $\nn$ such that $\| \nn - \rr \|_\infty = \max_{1 \le i \le 3} | n_i - r_i | \le L$ for some (small) radius $L>0$\footnote{For example, with $L=1$, $\mathcal{N}(\rr)$ represents a standard 27-connected neighbourhood of $\rr$.}. The term ``diffusion cube (DC) at $\rr$" will be used below to refer to a segment of DTI data {\it spatially} restricted to $\mathcal{N}_\rr$. Formally, the DC at $\rr$ is defined as
\[
s_\rr = s_\rr(\nn, \uu_k) : = \{s(\nn, \uu_k) \mid \nn \in \mathcal{N}_\rr, \, 1 \le k \le K \},
\]
where $s(\nn, \uu_k)$ denotes the signal value at position $\nn \in \mathcal{N}_\rr$ and diffusion-encoding orientation $\uu_k$. Thus, similarly to the entire dataset, $s_\rr$ can be viewed as a 4-D array of size $M \times M \times M \times K$, with $M = 2 L+1$.

\begin{figure}[t]
\begin{center}
\includegraphics[width=4.5in]{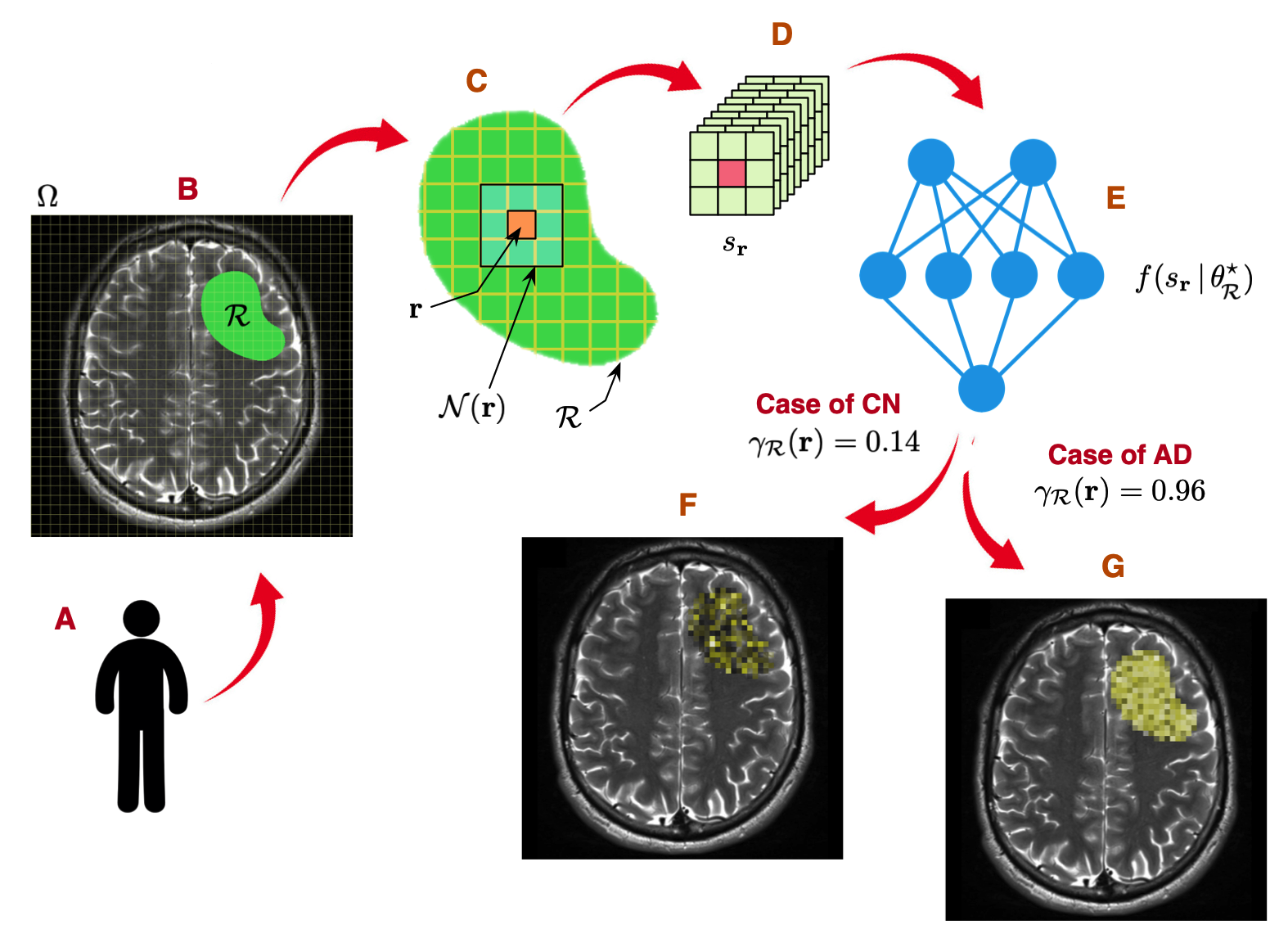}
\caption{Computation of PSIC: (A) subject under examination; (B) DTI dataset with a specified ROI $\mathcal{R} \in \Omega$; (C) ``inner" point $\rr$ and its neighbourhood $\NBR$; (D) corresponding DC $s_\rr$; (F) and (G) regional PSIC values in the case of CN and AD (G), respectively.}
\label{FGR1}
\end{center}
\end{figure}

As the next step, we refer to Fig.~\ref{FGR1} that illustrates various stages of computing the PSIC values in application to the stratification of CN and AD subjects. Suppose, for a given study subject (Subplot A); we are interested in making an inference based on DTI data confined to some prescribed ROI $\mathcal{R} \subset \Omega$ (Subplot B). Then, for each $\rr \in \mathcal{R}$, such that $\NBR \subset \mathcal{R}$ (Subplot C), its associated DC $s_\rr$ (Subplot D) is passed to a dedicated DNN $f(s_\rr \mid \theta)$ that yields a positive PSIC score $0 \le \gamma_\mathcal{R}(\rr) \le 1$. For the sake of argument, at the moment, the network parameters $\theta$ are assumed to be set to their optimal values $\theta^\ast_\mathcal{R}$. In this case, by virtue of network design and training, $\gamma_\mathcal{R}(\rr)$ is set up to scale proportionally to the likelihood of the neural tissue at $\rr \in \mathcal{R}$ to be affected by AD. Thus, in particular, the CN cases would be associated with relatively low values of PSIC (e.g., $\gamma_\mathcal{R}(\rr) = 0.14$), while, in the case of AD, the values would be in close vicinity of 1 (e.g., $\gamma_\mathcal{R}(\rr) = 0.96$).

Finally, the values of $\gamma_\mathcal{R}(\rr)$ computed at all $\rr \in \mathcal{R}$ can also be used as an imaging contrast, which can be superimposed over a structural display of underlying anatomy.  Such contrast is expected to be comparatively weak and scarce in CN (Subfigure F) while being intense and densely concentrated in AD (Subfigure G).
 
In practice, the optimal network parameters are estimated from training data. To this end, we hypothesize that, for a proper choice of $\mathcal{R}$, the effects of neurodegeneration are manifested in some hidden diffusion characteristics which persevere throughout the entire ROI. Moreover, such hidden characteristics are likely to be shared between different subjects within the same diagnostic group.

Under the above hypothesis, the training data can be formed by stockpiling the DCs {\it from all voxels} $\rr$ within a specified ROI {\it across all subjects} within the same diagnostic group. Specifically, in application to the binary problem of stratification of CN and AD subjects, such training data would consist of a set of DCs obtained from all available CN subjects, on the one hand, and a similar set of DCs coming from all available AD subjects, on the other hand.

\begin{figure}[t]
\begin{center}
\includegraphics[width=5in]{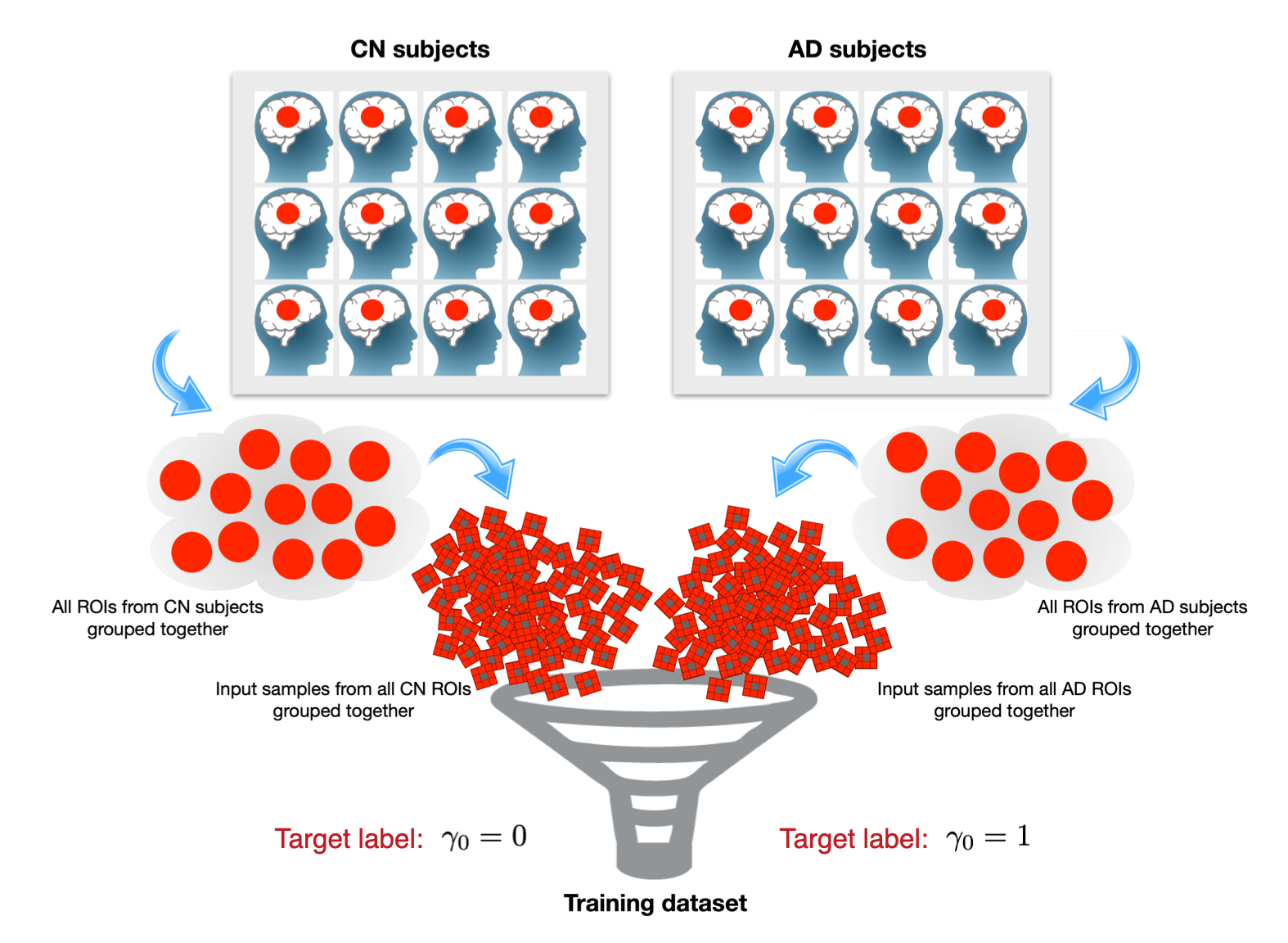}
\caption{Conceptual illustration of the process of formation of the training data for a specified ROI.}
\label{FGR2}
\end{center}
\end{figure}

More formally, let $N_{\rm CN}$ and $N_{\rm AD}$ be the number of subjects in the CN and AD groups, respectively. Also, let $\{ s^{{\rm{CN}}, i} \}_{i=1}^{N_{\rm CN}}$ (resp., $\{ s^{{\rm{AD}}, i} \}_{i=1}^{N_{\rm AD}}$) denote the diffusion datasets collected from the CN (resp., AD) subjects. In this case, the training data would consist of two subsets of samples defined as 
\[
\mathcal{X}_{CN}^\mathcal{R} := \left\{ \{ s_\rr^{{\rm{CN}}, i} \}_{\rr \in \mathcal{R}} \right\}_{i=1}^{N_{\rm CN}}, \quad
\mathcal{X}_{AD}^\mathcal{R} := \left\{ \{ s_\rr^{{\rm{AD}}, i} \}_{\rr \in \mathcal{R}} \right\}_{i=1}^{N_{\rm AD}},
\]  
along with their corresponding (target) labels $\gamma_0 = 0$ and $\gamma_0 = 1$, respectively. A conceptual illustration of the process of formation of the training data is shown in Fig.~\ref{FGR2}. 

In general, the above procedure can be applied to $M$ different ROIs $\{\mathcal{R}_j\}_{j=1}^M$, in which case one would have $M$ different training sets, $\{\mathcal{X}_{CN}^{\mathcal{R}_j}\}_{j=1}^M$ and $\{\mathcal{X}_{AD}^{\mathcal{R}_j}\}_{j=1}^M$. Each such set could then be used to estimate the optimal parameters $\theta^\star_{\mathcal{R}_j}$ for its related $\mathcal{R}_j$ independently.

% ---------------------------------------------------------------- SECTION 3 ------------------------------------------------------------------
\section{Analytical tools and principal operations}\label{Sec3}
\subsection{Composite Convolution}
At a conceptual level, the DNN proposed in this paper is based on a standard feed-forward configuration, consisting of a typical succession of convolutional operations interleaved with nonlinearities and resizing. However, when defining such operations, it would be amiss to disregard the physical properties of diffusion signals which offer a number of important advantages.

As discussed in Section~\ref{Sec2}, at its input, the network receives a single DC, which can be viewed as a 4-D numeric array of size $M \times M \times M \times K$. Alternatively, the DC could be considered to be an array of the discrete values of a diffusion signal $S(\xx,\uu)$ measured over $\Omega_L = \{ \nn \mid \| \nn \|_{\infty} \le L\}$ and spherical orientations $\{\uu_k\}_{k=1}^K$. Formally, $S(\xx,\uu)$ can be defined over the combined domain 
$\bar{\Omega}:={\rm Conv}(\Omega_L) \times \ES$, where ${\rm Conv}(\Omega_L)$ stands for the convex hull of $\Omega_L$. Thus, in order to incorporate convolution into the network design, one needs a proper definition of such operation for the signals defined over $\bar{\Omega}$.

Generally speaking, the convolution of $\bar{\Omega}$-domain signals is defined over the special Euclidean group $SE(3)$, working with which might be excessively complicated from a computational point of view. In what follows, we introduce a particular definition of this operation, which offers a number of important practical advantages. In this way, the present results expand on the simpler case of spherical signals over $\ES$, which have been successfully addressed in a number of recent studies \cite{Poulenard:2019aa,Mukhaimar:2020aa}.

The proposed method relies on a simplified interpretation of $SE(3)$ convolution, which is derived under the assumptions of {\it separability} and {\it zonality}. In particular, the former requires the convolution kernel to be a  separable product of a spatially-dependent and a spherically-dependent component. The assumption of zonality, on the other hand, requires the spherical component to be a {\it zonal} function, which implies its invariance to azimuthal rotations. Such functions admit representation in terms of Legendre polynomials $\{p_n\}_{n=0}^\infty$ as given by
\begin{equation}
\xi(t) = \sum_{n=0}^\infty \frac{2n+1}{4\pi} \, \xi_n \, p_n(t), \quad t \in [-1, +1],
\end{equation}
with $\xi_n = 2\pi \int_{-1}^1 \xi(t) p_n(t) \, dt$ known as the Legendre coefficient of degree $n$.

The use of zonal kernels considerably simplifies the definition of spherical convolution. To see that, we first assume that, at each $\xx \in \Omega_L$, $S(\xx, \uu)$ can be closely approximated by its truncated {\it Spherical Harmonic} (SH) expansion of the form
\begin{equation}\label{apprx}
S(\xx, \uu) \simeq \sum_{n=0, 2, \ldots}^\nmax \sum_{l=-n}^n c_{n, l}(\xx) Y_{n,l}(\uu),
\end{equation}
with $Y_{n,l}(\uu)$ and $c_{n,l}(\xx)$ being the $l$-th order SH of degree $n$ and its corresponding expansion coefficient, respectively. Note that, due to the spherical symmetry of DTI signals, the summation in \eqref{apprx} is restricted to the even values of $n$, in which case the total number of expansion coefficients is equal to $P= 0.5 \, (\nmax+1)(\nmax+2)$, for a predefined maximal degree $\nmax > 0$.

At any $\xx \in \Omega_L$, the availability of the SH coefficients $c_{n,l}(\xx)$ and the knowledge of the Legendre coefficients $\xi_n$ of the zonal kernel $\xi(\uu)$ can be used to define the spherical convolution of $S(\xx, \cdot)$ and $\xi$ according to
\begin{equation}\label{Sconv}
\left( S(\xx, \cdot) \ast_{\uu} \xi \right) (\uu)= \sum_{n=0, 2, \ldots}^\nmax \sum_{l=-n}^n \tilde{c}_{n, l}(\xx) Y_{n,l}(\uu),
\end{equation}
where
\[
\tilde{c}_{n, l}(\xx) = \xi_n \, c_{n, l}(\xx),
\]
for all $n = 0, 2, \ldots, \nmax$ and $|l| \le n$.

Next, we note that at each $\xx \in \Omega_L$, $c_{n, l}(\xx)$ is just a real vector of length $P$. When computed over the entire $\Omega_L$; all such vectors can be conveniently assembled into a 4-D array $c$ of size $M \times M \times M \times P$. Alternatively, $c$ can be viewed as a collection of $P$ volumetric ``coefficient images" $c_{n,l} \in \mathbb{R}^{M \times M \times M}$, where each $c_{n,l}$ comprises the spatially-dependent values of the SH coefficient of degree $n$ and order $l$.

It is also convenient to split $c$ into $(\nmax + 2)/2$ subgroups $c_k$ of 3-D arrays according to the value of their degree $n$, i.e., $c_k = \{ c_{k,l} \}_{l=-k}^k$, with $k=0, 2, \ldots, \nmax$. In this case, the spherical convolution in \eqref{Sconv} amounts to scaling each ``coefficient image" in $c_n$ by the {\it same} scalar $\xi_n$. While computationally efficient, however, this operation has the disadvantage of ignoring the spatial behaviour of input signals.

The coordinate-wise spherical convolution in \eqref{Sconv} can be generalized into a {\it composite} spatial-spherical convolution as follows. For $n \le \nmax$, let $w_n = \{w_{n,l}^k\}_{|l|\le n, |k|\le n}$ be a set of $P_n^2$ spatial-domain filters of size $J \times J \times J$. Then, based on the assumption of separability, the operation of $n^{\rm th}$-{\it band spatio-spherical filtering} can be defined as
\begin{equation}\label{ConvFinal}
\hat{c}_{n, l} = \sum_{k=-n}^n c_{n,k} \ast_\rr w_{n, l}^k, \quad \mbox{for all} \,\, |l| \le n,
\end{equation}
where $\ast_\rr$ stands for discrete convolution in the spatial domain. The definition in \eqref{ConvFinal} suggests that, for each spherical order $l$, the output $\hat{c}_{n,l}$ is computed as a linear {\it convolutional} combination of the ``input images" in $c_n$ with filters $w_{n,l}^{-n}$, ..., $w_{n,l}^{0}$, ... , $w_{n,l}^{n}$. In this way, the action of \eqref{ConvFinal} extends across the entire $\Omega_L$, as opposed to the case of \eqref{Sconv}.

Extending the $n^{\rm th}$-band spatio-spherical filtering in \eqref{ConvFinal} to all $n = 0, 2, \ldots, \nmax$ gives rise to a convolution-type operator that takes effect across both of the spherical and spatial domains. In what follows, this operation will be referred below to as {\it composite convolution}\footnote{For each $n$, the computation of $\tilde{c}_n$ can be identified with the action of multi-channel convolution that is a standard computational routine included in many existing DL frameworks, such as, e.g., TensorFlow\textregistered{} (which has been used in this study).}. Formally, for a 4-D array of SH coefficients $c$ and a bank of filters $w = \{w_0, w_2, \ldots, w_\nmax\}$, their composite convolution
\begin{equation}\label{ConvComp}
\tilde{c} = c \ast_{\uu, \rr} w
\end{equation}
is defined according to \eqref{ConvFinal} for each $n = 0, 2, \ldots, \nmax$ and $|l| \le n$.

\subsection{SH coefficients}
The composite convolutional in \eqref{ConvComp} implies the availability of SH coefficients $c$, which can be estimated directly from DTI data, as described, e.g., in \cite{Descoteaux:2006}. The estimation is carried out independently on each of the $M^3$ signals comprising a given DC, resulting in an $M \times M \times M \times P$ array of associated SH coefficients $c$. The proposed DNN has been designed to work directly with such $c$, which are assumed to be precomputed prior to network training. For the sake of  notational simplicity, such input arrays will also be referred bellow to as ``DCs".

Finally, the question of setting $\nmax$ should not be overlooked. Usually, $\nmax$ is defined in accordance with a required $b$-value. In particular, for $b = 1000$ s/mm$^2$ (as used in ADNI-II), setting $\nmax = 6$ seems to be a conventional choice \cite{Rohde:2004,Jones:2004aa,Alexander:2007}. Note that, in this case, the total number of SHs is equal to $P = 28$.

% ---------------------------------------------------------------- SECTION 4 ------------------------------------------------------------------
\section{Proposed network architecture}\label{Sec4}
\subsection{Convolutional layers}
The proposed DNN consists of several convolutional layers, each of which is parameterized by (convolutional) {\it weights} $w = \{w_n\}_{n=0,2,\ldots}$ and a vector of $P$ (scalar) {\it biases} $b = \{b_{n,l}\}$, with $n =0, 2, \ldots, \nmax$ and $|l| \le n$. Given an input array $c^{\rm in}$, each such layer computes its output $c^{\rm out}$ according to
\begin{equation}\label{ConvLay}
c^{\rm out} = \left( c^{\rm in} \ast_{\uu, \rr} w \right) + b, 
\end{equation}
where the plus sign is assumed to broadcast the values of $b$ so that every ``coefficient image" in $(c^{\rm in} \ast_{\uu, \rr} w)$ is summed with a different $b_{n,l}$. Note that the affine operation in \eqref{ConvLay} is not exclusive to volumetric data, since it can be reduced to its 2-D and 1-D versions by merely replacing all $w_{n,l}^k$ in $w$ by their 2-D and 1-D counterparts, respectively.

\subsection{Activation} The scope of activation functions currently used in DL is broad. In this work, all activation functions have had the form of a basic {\em rectified linear unit} (ReLU) \cite{LeCun:2015aa}, with its input-output relation defined as given by
\begin{equation}\label{ReLu}
c^{\rm out} = {\rm ReLU}(c^{\rm in}) = \max ( 0,  c^{\rm in}),
\end{equation}
where the maximization is carried out independently for all values in the input array. Note that, over the last years, \eqref{ReLu} has been modified in several ways (resulting in, e.g., leaky ReLu, noisy ReLu, and exponential ReLU). In our experiments, however, the basic definition in \eqref{ReLu} was observed to work more than adequately.

\subsection{Pooling}\label{POOL}
Pooling is a standard method of data aggregation which can be used to suppress redundancies in input data, reduce the number of network parameters, and to minimize the risk of overfitting \cite{Goodfellow:2016aa}. The special structure of DC samples $c$, however, requires a proper adaptation of this operation. Specifically, in this work, the operation of pooling consisted of {\it max pooling} applied along a {\it singular spatial dimension}. Specifically, depending on the direction of maximization, the pooling can be defined in three possible ways as given by
\begin{subequations}\label{P3}
\begin{align}
\cout(y, z) = \mathcal{P}_3^x(\cin(x, y, z)) &:=  \max_{x}  \cin(x, y, z), \\
\cout(x, z) = \mathcal{P}_3^y(\cin(x, y, z)) &:=  \max_{y}  \cin(x, y, z), \\
\cout(x, y) = \mathcal{P}_3^z(\cin(x, y, z)) &:=  \max_{z}  \cin(x, y, z),
\end{align}
\end{subequations}
for each $(n, l)$ and $(x,y,z) \in \Omega_L$. Above, $\mathcal{P}$ denotes the pooling operator, with its sub- and superscripts indicating the spatial dimensionality of $\cin$ and the direction of maximization, respectively.

It is important to emphasize, while $\cin$ in \eqref{P3} depends on three spatial variables, the number of spatial dimensions of $\cout$ is reduced to two, and, consequently, each of the $P$ ``coefficient images" composing $c^{\rm out}$ is now an array of size $M \times M \times P$. The resulting outputs could be subjected to 2-D pooling defined according to 
\begin{subequations}\label{P2}
\begin{align}
\cout(y) = \mathcal{P}_2^{x}(\cin(x, y) = \max_{x}  \cin(x, y), \\
\cout(x) = \mathcal{P}_2^{y}(\cin(x, y)) = \max_{i_y'}  \cin(x, y),
\end{align}
\end{subequations}
for each $(n,l)$ and $(x, y)$, with $|x|, |y| \le L$. Note that, in both variants above, the spatial dimension of $c^{\rm out}$ has been reduced to one (i.e., each $\cout$ is now an array of size $M \times P$).

Proceeding analogously, one can finally define the operation of 1-D pooling as
\begin{equation}\label{P1}
\cout = \mathcal{P}_1^{x}(\cin(x)) =  \max_{x}  \cin(x),
\end{equation}
for each $(n,l)$ and $|x| \le L$. This type of pooling collapses the spatial dimension of $c^{\rm in}$, resulting in a length-$P$ vector $c^{\rm out}$ of modified SH coefficients.

\begin{figure}[t]
\begin{center}
\hspace*{-1.1in} 
\includegraphics[width=6.5in]{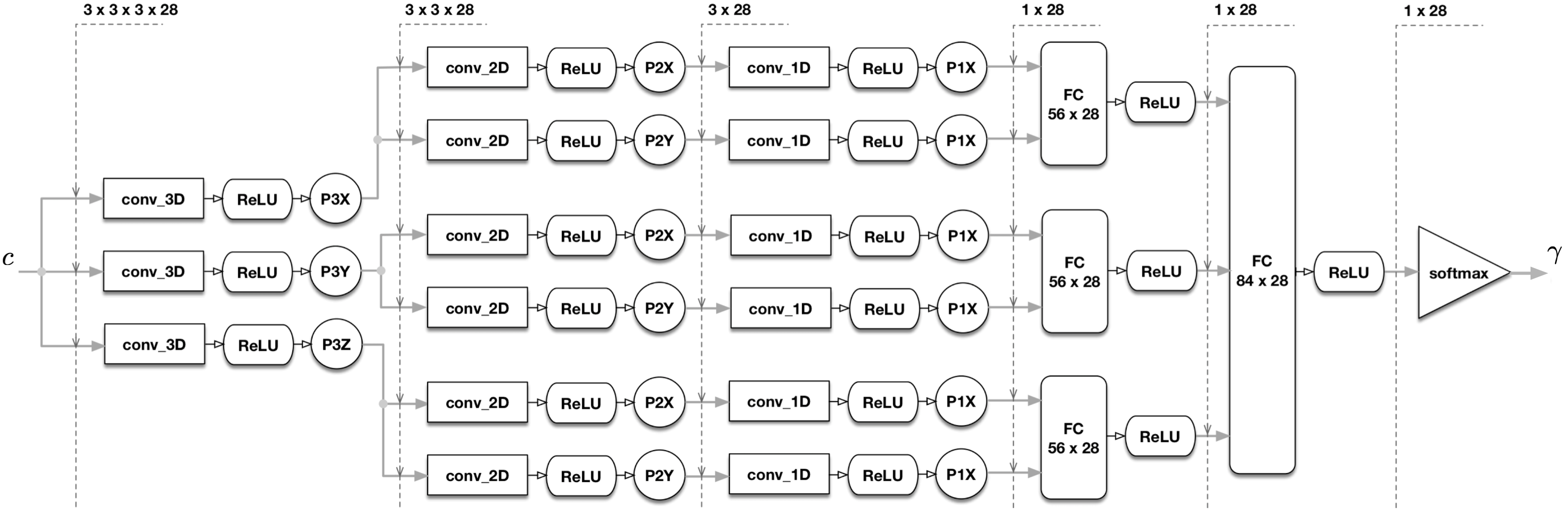}
\caption{Proposed network architecture.}
\label{F1}
\end{center}
\end{figure}

\subsection{Network Architecture} 
Apart from using the purpose-built operations of convolution and pooling, the proposed DNN relies on a standard feed-forward architecture, which is depicted in Fig.~\ref{F1} for the case of $L=1$ and $\nmax=6$ (i.e., $M=3$ and $P = 28$). The network consists of six neural layers which, in the figure, are shown separated by vertical dotted lines. The dimensions specified at the top of these lines indicate the size of the inputs received by each layer. Specifically, the input layer of the network receives a DC array $c$ of size $3 \times 3 \times 3 \times 28$, which is subjected to three different transformations of type \eqref{ConvLay}, followed by ReLU activation and pooling along the three spatial coordinately. Thus, the second layer of the network receives a total of three arrays of size $3 \times 3 \times 28$, each of which is processed in an analogous manner, using \eqref{P2} and \eqref{ConvLay} (with a 2-D version of \eqref{ConvComp}).

Similar computations are applied to all of the six inputs of the third layer, which are subjected to transformation \eqref{ConvLay} (with a 1-D version of \eqref{ConvComp}), ReLU activation, and 1-D pooling according to \eqref{P1}. Subsequently, each of the resulting pairs of $P$-length vectors is fused into a single $28$-length vector by means of a {\it fully connected layer} (FCL). An additional FCL is used to transform its inputs into a single vector of length $28$. 

The final layer of the DNN consists of a linear transformation into the space of diagnostic outcomes, followed by the {\it softmax} operator. In the case of binary stratification (i.e., CN vs AD), the latter can be defined as  \cite{Goodfellow:2016aa}
\begin{equation}
\gamma = \frac{e^\alpha}{e^\alpha + e^\beta},
\end{equation} 
with $\alpha$ and $\beta$ being the two outputs of the linear transformation. In the context of this paper, the score $0 \le \gamma \le 1$, thus computed, is referred to as PSIC.

\subsection{Network Optimization} The proposed DNN is parameterized by the values of convolution kernels, biases and weight matrices used across various components of the network. Altogether, these parameters can be gathered into a single vector $\theta$, in which case, for a given $\theta$, the DNN acts as a forward mapping $\gamma = f(c \mid \theta)$, associating each DC $c$ with its PSIC score. In this case, it is a standard practice to estimate the optimal value of $\theta$ via solving a cross-entropy minimization problem of the form
\begin{equation}\label{optim}
\theta^\ast = \arg\min_\theta \mathcal{E}_c \left\{ -\gamma_0 \log f(c \mid \theta) \right\},
\end{equation}
where expectation $\mathcal{E}$ is computed over the empirical distribution of $c$.

Finally, let $\{\mathcal{R}_j\}_{j=1}^M$ be a set of relevant ROIs. Also, for each $\mathcal{R}_j$, let $\theta^\ast_{\mathcal{R}_j}$ denote the optimal values of its associated DNN parameters. Then, given an unlabelled set of DTI measurements, $f(\cdot \mid \theta^\ast_{\mathcal{R}_j})$ can be used to compute the PSIC values across the entire $\mathcal{R}_j$. Subsequently, the resulting scores can be summarized into regional statistics for the purpose of subject stratification, as described next.

% ---------------------------------------------------------------- SECTION 5 ------------------------------------------------------------------
\section{Experimental study design}\label{Sec5}
\subsection{Study data} As stated earlier, the experimental study of this work has been focused on the problem of stratification of CN and AD subjects. To this end, 20 CN and 20 AD age-matched subjects (mean age 72.6 $\pm$ 7.6 years) were selected from the DTI database of ADNI-II\footnote{For more details on the inclusion/exclusion criteria and other design parameters used by the ADNI-II study, please refer to {\tt adni.loni.usc.edu}.}. The dataset of each subject consisted of $K=41$ diffusion-encoded volumetric scans, acquired at $b=1000$ s/mm$^2$, as well as five $b_0$-volumes (i.e., scans acquired in the absence of diffusion sensitization). In addition, each dataset was supplemented with its associated ${\rm T}_1$- and ${\rm T}_2$-weighted scans required for structural image alignment and segmentation.

\subsection{Data preprocessing} For each dataset, its $b_0$-volumes were first co-registered and then merged into a single (average) $b_0$-volume, which was subsequently used for normalization of the $K$ diffusion-encoded volumes. After that, the normalized data were subjected to preprocessing by means of a custom pipeline which had been designed based on the recommendations of \cite{Glasser:2013aa}. The technical implementation of the pipeline relied on the NiPype framework of NiPy$\textregistered$ ({\tt nipype.readthedocs.io}), which allowed convenient integration of many well-established tools of computational imaging, including FSL ({\tt https://fsl.fmrib.ox.ac.uk/}), ANTs ({\tt picsl.} {\tt upenn.edu/software/ants/}), and SPM ({\tt www.fil.ion.ucl.ac.uk/spm/}). As per usual, the principal purpose of the preprocessing pipeline has been to compensate for various imaging artefacts caused by the effects of subject motion, variable magnetic susceptibility, eddy currents, etc.

Additionally, FreeSurfer \cite{FS:2012aa} ({\tt surfer.nmr.mgh.harvard.edu/}) was used for the purpose of segmentation of grey and white matter, with their subsequent parcellation into smaller anatomical regions. Subsequently, the anatomic labels provided by FreeSurfer were used to partition the image domain $\Omega$ of the DTI volumes into a set of predefined ROIs $\{\mathcal{R}_j\}_{j=1}^M$, as detailed below.

\subsection{Definition of ROIs} The two left columns of Table~\ref{T1} summarize the names of the anatomical regions computed by FreeSurfer, along with their acronyms. In addition, for each ROI, the leftmost columns of the table indicate the total number of DC samples available in its respective CN ($\# \mathcal{X}_{\rm CN}^{\mathcal{R}_j}$) and AD ($\# \mathcal{X}_{\rm AD}^{\mathcal{R}_j}$) subsets. 

The ROIs have been chosen to correspond to different parts of white matter anatomy, which are known to be implicated in the pathogenesis of AD. It should be noted, however, that FreeSurfer lacks means of {\it direct} delineation of white matter. Instead, various elements of the latter are labelled based on their proximity to the nearby anatomical structures of grey matter. Thus, for instance, all voxels designated as white matter and located, e.g., within a 5 mm ribbon around the left superior frontal gyrus would be labelled as lh-SFW. This approach is obviously not without limitations, the rectification of which has been in the focus of ongoing research \cite{Huo:2018aa}. Nevertheless, considering the comparative nature of the results reported here, the choice of a specific method of whole-brain segmentation does not seem to be particularly critical.

\begingroup
\begin{table}[h]
\centering
\begin{tabular}{@{} cllccc @{}}
\toprule
{$j$} & {\bf ROI name} & {\bf Acronym} & {$\# \mathcal{X}_{\rm AD}^{\mathcal{R}_j}$} & {$\# \mathcal{X}_{\rm CN}^{\mathcal{R}_j}$} & {\bf Total} \\
\midrule
%Cerebellum cortex (lh) 			& lh-CBC  	& 38350	& 35139	& 73489 \\
%Cerebellum cortex (rh) 			& rh-CBC		& 40805	& 41264 	& 82069 \\
1 & Superior frontal wm (lh)			& lh-SFW		& 17251	& 15935 	& 33186 \\
2 & Superior frontal wm (rh) 			& rh-SFW		& 16768	& 16739 	& 33507 \\
3 & Cerebellum wm (lh)				& lh-CBW		& 16010	& 13686 	& 29696 \\
4 & Cerebellum wm (rh)				& rh-CBW		& 16510	& 14397 	& 30907 \\
5 & Precentral wm (lh)				& lh-PreCW	& 10028	& 8453 	& 18481 \\
6 & Precentral wm (rh)				& rh-PreCW	& 9539 	& 8147 	& 17686 \\
7 & Inferior parietal wm (lh) 			& lh-IPW		& 4442 	& 3841 	& 8283 \\
8 & Inferior parietal wm (rh) 			& rh-IPW		& 5452	& 4678 	& 10130 \\
9 & Middle frontal wm, rostral (lh)		& lh-MFWros	& 4397	& 4593	& 8990 \\
10 & Middle frontal wm, rostral	(rh)		& rh-MFWros	& 3809	& 4027	& 7836 \\
11 & Precuneus wm (lh) 				& lh-PCUNW	& 3046	& 2970 	& 6016 \\
12 & Precuneus wm (rh)	 			& rh-PCUNW	& 3540	& 3416 	& 6956 \\
13 & Superior parietal wm (lh)	 		& lh-SPW		& 3184	& 3010 	& 6194 \\
14 & Superior parietal wm (rh)			& rh-SPW		& 2989	& 3139 	& 6128 \\
\bottomrule
\end{tabular}
\caption{Selected ROIs and their associated number of input samples in $\mathcal{X}_{\rm CN}^{\mathcal{R}_j}$ and $\mathcal{X}_{\rm AD}^{\mathcal{R}_j}$ (shown in the three rightmost columns). Note that the abbreviations ``wm", ``lh", and ``rh" stand for white matter, left and right hemisphere, respectively.}
\label{T1}
\end{table}
\endgroup

Using the methodology of Section~\ref{Sec2}, each of the $M=14$ regions in Table~\ref{T1} was used to assemble its associated input samples $\mathcal{X}_{\rm CN}^{\mathcal{R}_j}$ and $\mathcal{X}_{\rm AD}^{\mathcal{R}_j}$, corresponding to the target labels $\gamma_0=0$ and $\gamma_0=1$, respectively. Subsequently, the process of network training was carried out for each $j$ independently, yielding a vector of optimal network parameters for each of the chosen ROIs.

\subsection{Network training} Prior to training, for each $j = 1, 2, \ldots, 14$, the samples in $\mathcal{X}_{\rm CN}^{\mathcal{R}_j}$ and $\mathcal{X}_{\rm AD}^{\mathcal{R}_j}$ were randomized and split into a {\it training} and a {\it validation} dataset (in proportion 4:1), which were subsequently used for the purpose of estimation of $\theta^\ast_{\mathcal{R}_j}$ and final performance evaluation, respectively. In all cases, the optimization was performed by means of the adaptive moment estimation (Adam) algorithm \cite{ADAM}, with a fixed learning rate of $0.5 \cdot 10^{-3}$,  batch size of 256 samples, and 200 epochs. The network training procedure was augmented with dropout regularization, with the value of keep probability set to 0.7 to alleviate the effects of overfitting.

The convergence of optimization has been monitored in terms of empirical {\em prediction accuracy} (PA). Given a set of $N_L$ labelled DC samples $\{c_l, \gamma_l\}_{l=1}^{N_L}$, with $\gamma_l \in \{0, 1\}$, the latter can be defined as   
\begin{equation}\label{PA}
{\rm PA}(\theta) = 1- \frac{1}{N_L} \sum_{l =1}^{N_L} | \hat{\gamma}_l(c_{l} \,|\, \theta) - \gamma_l |,
\end{equation}
where $\hat{\gamma}_l(c_{l} \,|\, \theta) = 1$, if $f(c_{l} \,|\, \theta) \ge 0.5$ and $\hat{\gamma}_l(c_{l} \,|\, \theta) = 0$, otherwise.

In the context of network training, the PA criterion can be a useful indication of the convergence of stochastic gradient. On the other hand, when used on validation data, PA provides a more objective measure of both optimality and {\it generalizability} in view of its virtual independence of the effects of overfitting. Thus, the higher values of validation PA are typically indicative of more accurate performance, in general \cite{Goodfellow:2016aa}.

\subsection{Binary classification} Once the training is complete, and the values of $\theta^\ast_{\mathcal{R}_j}$ are available for each ROI $\mathcal{R}_j$, the optimal forward mappings $f(\cdot \mid \theta^\ast_{\mathcal{R}_j})$ can be used for stratification of unclassified subjects. In particular, an {\it unlabelled} dataset of DTI measurements can be converted into $M$ subsets of DC samples $\{c_\rr\}_{\rr \in \mathcal{R}_j}$ in accordance with the selected ROIs. Subsequently, for each $\mathcal{R}_j$, its respective DC samples can be converted into a set of PSIC scores $\{\gamma_\rr\}_{\rr \in \mathcal{R}_j}$, with $\gamma_\rr = f(c_\rr \mid \theta^\ast_{\mathcal{R}_j})$. In this case, the subject could be stratified based on
\begin{equation}\label{NPC0}
\underset{\rr \in \mathcal{R}_j}{\rm median}(\gamma_\rr) \begin{array}{c} {\scriptstyle{\rm AD}} \\ \gtrless \\ {\scriptstyle{\rm CN}} \end{array} 0.5, 
\end{equation}
with the decision made independently for each $j$. In such case, the median PSIC score on the left side of \eqref{NPC0} can be viewed as a cumulative regional marker of AD.

Furthermore, for the same subject, the PSIC values in $\{\{\gamma_\rr\}_{\rr \in \mathcal{R}_j}\}_{j=1}^M$ can serve in the role of imaging contrast. Particularly, these values can be superimposed on structural MRI scans, thus offering the possibility of {\it visual} exploration of the spatial variability of PSIC values.

\subsection{Reference methods} 
The performance of the proposed classifier has been compared against region-based classification based on multiple diffusion metrics and their combination. The selected metrics included four standard DTI measures, {\it viz.}: {\it mean diffusivity} (MD), {\it fractional anisotropy} (FA), as well as diffusion {\it linearity (CL)} and {\it planarity} \cite{Mori:2014aa}. The list of metrics also included {\it diffusion volume} (DV), {\it average sample diffusion} (ASD), {\it diffusion energy} (DE), and the {\it coefficient of variation of diffusion} (CVD), which can provide more general characterization of diffusion dynamics, independent of DTI modelling \cite{Aja-Fernandez:2018aa}. It goes without saying, the above list is by no means exclusive, and other useful characteristics of cerebral diffusion could have been included as well \cite{Novikov:2018aa}. This being the case, however, besides covering both basic and advanced options, the selected metrics have had an important advantage of {\it estimability} based on relatively small datasets, as it is the case in the present study.

In this paper, the region-based classification was based on the likelihood ratio test \cite{Rigollet:2011aa}. Specifically, for each metric $\mu_i$, with $i=1,2,\ldots, 8$, its values inside $\mathcal{R}_j$ were assumed to be independent realizations of a random variable, with its probability densities in the CN and AD subgroups given by $p^{\rm CN}_{\mathcal{R}_j}(\mu_i)$ and $p^{\rm AD}_{\mathcal{R}_j}(\mu_i)$, respectively. Then, given an unlabelled dataset, the observed values of $\{\mu_i(\rr)\}_{\rr \in \mathcal{R}_j}$ were used to stratify the subject according to
\begin{equation}\label{NPC2}
\frac{\sum_{\rr \in \mathcal{R}_j} \log p_{\mathcal{R}_j}^{AD}(\mu_i(\rr))}{\sum_{\rr \in \mathcal{R}_j} \log p_{\mathcal{R}_j}^{CN}(\mu_i(\rr))} 
\begin{array}{c} {\scriptstyle{\rm AD}} \\ \gtrless \\ {\scriptstyle{\rm CN}} \end{array} \eta,
\end{equation} 
for some decision variable $\eta$. For each $\mu_i$ and $\mathcal{R}_j$, $\eta$ had been set to its optimal value through maximizing the area under the {\it receiver operating characteristic} (ROC) curve of its corresponding classifier. In each case, the probability densities in \eqref{NPC2} were assumed to be Gaussian, with their means and variances estimated from the available data, following a standard leave-one-out cross-validation procedure \cite{Konishi:1992aa}.

An addition reference method was based on concurrent use of multiple diffusion metrics within the framework of  {\it logistic regression} (LR). Similarly to the proposed DNN-based classifier, LR was used to map the input values of $\{\mu_i\}_{i=1}^8$ into a scalar classification score \cite{Agresti:2002aa}.

% ---------------------------------------------------------------- SECTION 6 ------------------------------------------------------------------
\section{Experimental results}\label{Sec6}
The main experimental results of this study are summarized in Tables \ref{T2} and \ref{T3}, where the former shows the PA scores produced by different classifiers for different ROIs (with the proposed method of classification denoted by ``DNN"). One can see that the DNN-based classifier considerably outperforms the reference approaches across all ROIs. There is, however, a slight decrease in DNN's performance for $\mathcal{R}_j$ with $j > 10$, which is likely due to the reduction in the total number of training samples available for these ROIs (as indicated in Table~\ref{T1}).

As evidenced by Table~\ref{T2}, among the reference methods, the LR classifier showed superior performance in less than half of the cases. In other cases, the accuracy of classification was observed to depend on a particular metric/ROI combination. Interestingly enough, in about half of such cases, the most basic DTI metrics (such as MD and FA) demonstrated better performance in comparison to more advanced options.

\begin{table}[tp]
\caption{PA values produced by different classifiers for various $\mathcal{R}_j$. Note that, for each $j$, the two best results are outlined in bold.}
\begin{center}
\begin{tabular}{| c | l || c | c | c | c | c | c | c | c | c | c |}
\hline
$j$ & {\bf ROI, $\mathcal{R}_j$} & MD & FA & CL & CP & DV & ASD & DE & CVD & LR & {\bf DNN} \\
 \hline
 \hline
% & lh-CBC	  	& 0.59 & {\bf 0.64} & 0.62 & 0.51 & 0.59 & 0.59 & 0.59 & 0.56 & 0.59 & {\bf 0.97} \\
% & rh-CBC		& 0.62 & {\bf 0.74} & 0.72 & 0.59 & 0.56 & 0.62 & 0.51 & 0.56 & 0.69 & {\bf 1.00} \\
1 & lh-SFW		& 0.69 & 0.62 & {\bf 0.72} & 0.59 & 0.67 & 0.69 & 0.67 & 0.54 & {\bf 0.72} & {\bf 0.97} \\
2 & rh-SFW		& 0.69 & 0.54 & 0.54	 & 0.59 & 0.69 & 0.69 & 0.67 & 0.49 & {\bf 0.72} & {\bf 0.97} \\
3 & lh-CBW		& 0.46 & 0.64 & 0.64	 & 0.64 & 0.54 & 0.46 & 0.49 & 0.64 & {\bf 0.72} & {\bf 1.00} \\
4 & rh-CBW		& 0.54 & 0.64 & {\bf 0.77} & 0.51 & 0.54 & 0.54 & 0.54 & 0.69 & 0.62 & {\bf 1.00} \\
5 & lh-PreCW		& 0.69 & {\bf 0.82} & 0.79 & 0.51 & 0.69 & 0.69 & 0.69 & 0.72 & {\bf 0.82} & {\bf 1.00} \\
6 & rh-PreCW		& {\bf 0.69} & 0.67 & 0.67 & 0.62 & 0.72 & {\bf 0.69} & 0.64 & 0.62 & 0.67 & {\bf 0.97} \\
7 & lh-IPW		& 0.74 & 0.67 & 0.74	 & 0.59 & {\bf 0.77} & 0.74 & {\bf 0.77} & 0.64 & 0.72 & {\bf 0.95} \\
8 & rh-IPW		& 0.79 & 0.74 & 0.74	 & 0.64 & 0.79 & 0.79 & {\bf 0.82} & 0.69 & 0.77 & {\bf 0.97} \\
9 & lh-MFWros	& 0.72 & 0.64 & 0.74 & 0.59 & 0.69 & 0.72 & {\bf 0.77} & 0.62 & 0.59 & {\bf 0.97} \\
10 & rh-MFWros	& {\bf 0.69} & 0.59 & 0.49	& 0.64 & {\bf 0.69} & {\bf 0.69} & 0.67 & 0.54 & 0.67 & {\bf 1.00} \\
11 & lh-PCUNW	& 0.72 & 0.59 & 0.59 & 0.44 & {\bf 0.74} & 0.72 & {\bf 0.74} & 0.44 & 0.56 & {\bf 0.92} \\
12 & rh-PCUNW	& 0.79 & 0.54 & 0.49	 & 0.46 & 0.82 & 0.79 & {\bf 0.85} & 0.54 & 0.62 & {\bf 0.95} \\
13 & lh-SPW		& 0.56 & 0.64 & 0.62	 & 0.49 & 0.54 & 0.56 & 0.56 & 0.54 & {\bf 0.72} & {\bf 0.87} \\
14 & rh-SPW		& {\bf 0.77} & 0.67 & 0.69	& 0.56 & 0.74 & {\bf 0.77} & 0.72 & 0.67 & {\bf 0.77} & {\bf 0.92} \\
\hline
\end{tabular}
\end{center}
\label{T2}
\end{table} 

\begin{table}[tp]
\caption{AUC values produced by different classifiers for various $\mathcal{R}_j$. Note that, for each $j$, the two best results are outlined in bold.}
\begin{center}
\begin{tabular}{| c | l || c | c | c | c | c | c | c | c | c | c |}
\hline
$j$ & {\bf ROI, $\mathcal{R}_j$} & MD & FA & CL & CP & DV & ASD & DE & CVD & LR & {\bf DNN} \\
 \hline
 \hline
% & lh-CBC	  	& 0.66 & 0.29 & 0.27 & 0.41 & 0.67 & 0.66 & 0.66 & 0.37 & {\bf 0.74} & {\bf 1.00} \\
% & rh-CBC	  	& 0.63 & 0.28 & 0.25 & 0.46 & 0.63 & 0.63 & 0.62 & 0.41 & {\bf 0.76} & {\bf 1.00} \\
1 & lh-SFW		& 0.73 & 0.37 & 0.31 & 0.61 & 0.73 & 0.73 & 0.72 & 0.46 & {\bf 0.83} & {\bf 0.97} \\
2 & rh-SFW		& {\bf 0.74} & 0.38 & 0.39 & 0.45 & {\bf 0.74} & {\bf 0.74} & 0.73 & 0.45 & {\bf 0.74} & {\bf 0.98} \\
3 & lh-CBW		& 0.54 & 0.36 & 0.39 & 0.31 & 0.54 & 0.54 & 0.53 & 0.39 & {\bf 0.74} & {\bf 1.00} \\
4 & rh-CBW		& 0.46 & 0.31 & 0.30 & 0.48 & 0.44 & 0.46 & 0.43 & 0.31 & {\bf 0.65} & {\bf 1.00} \\
5 & lh-PreCW		& 0.82 & 0.19 & 0.18 & 0.44 & 0.82 & 0.82 & 0.80 & 0.27 & {\bf 0.89} & {\bf 1.00} \\
6 & rh-PreCW		& 0.72 & 0.28 & 0.26 & 0.47 & 0.72 & 0.72 & 0.72 & 0.34 & {\bf 0.79} & {\bf 0.99} \\
7 & lh-IPW		& 0.84 & 0.31 & 0.21 & 0.55 & 0.84 & 0.84 & 0.84 & 0.39 & {\bf 0.85} & {\bf 0.98} \\
8 & rh-IPW		& 0.84 & 0.29 & 0.21 & 0.44 & {\bf 0.86} & 0.84 & 0.87 & 0.36 & 0.84 & {\bf 0.99} \\
9 & lh-MFWros	& {\bf 0.76} & 0.32 & 0.33 & 0.47 & {\bf 0.76} & {\bf 0.76} & 0.75 & 0.36 & 0.69 & {\bf 0.97} \\
10 & rh-MFWros	& {\bf 0.74} & 0.47 & 0.54 & 0.38 & {\bf 0.74} & {\bf 0.74} & {\bf 0.74} & 0.47 & 0.71 & {\bf 1.00} \\
11 & lh-PCUNW	& 0.81 & 0.41 & 0.38 & 0.49 & 0.83 & 0.81 & {\bf 0.84} & 0.48 & 0.63 & {\bf 0.98} \\
12 & rh-PCUNW	& 0.85 & 0.45 & 0.42 & 0.62 & {\bf 0.87} & 0.84 & {\bf 0.87} & 0.50 & 0.57 & {\bf 0.97} \\
13 & lh-SPW		& 0.74 & 0.36 & 0.31 & 0.44 & 0.74 & 0.74 & 0.73 & 0.40 & {\bf 0.77} & {\bf 0.93} \\
14 & rh-SPW		& 0.82 & 0.27 & 0.23 & 0.54 & 0.82 & 0.82 & 0.82 & 0.31 & {\bf 0.83} & {\bf 0.94} \\
\hline
\end{tabular}
\end{center}
\label{T3}
\end{table}

In addition, leave-one-out cross-validation \cite{Konishi:1992aa} was used to compare the proposed and reference classifiers in terms of their respective ROC curves. Specifically, the optimality of classification was assessed in terms of the {\it area-under-curve} (AUC) criterion, whose values are shown in Table~\ref{T3}. Here, the higher values of AUC indicate a higher accuracy of classification, with the upper bound of 1 corresponding to the case of perfect classification. Thus, as demonstrated by the table, the proposed classifier offers a notably better solution to the problem of CN/AD stratification in comparison with the reference ones.

\begin{figure}[!t]
\begin{center}
\includegraphics[width=5in]{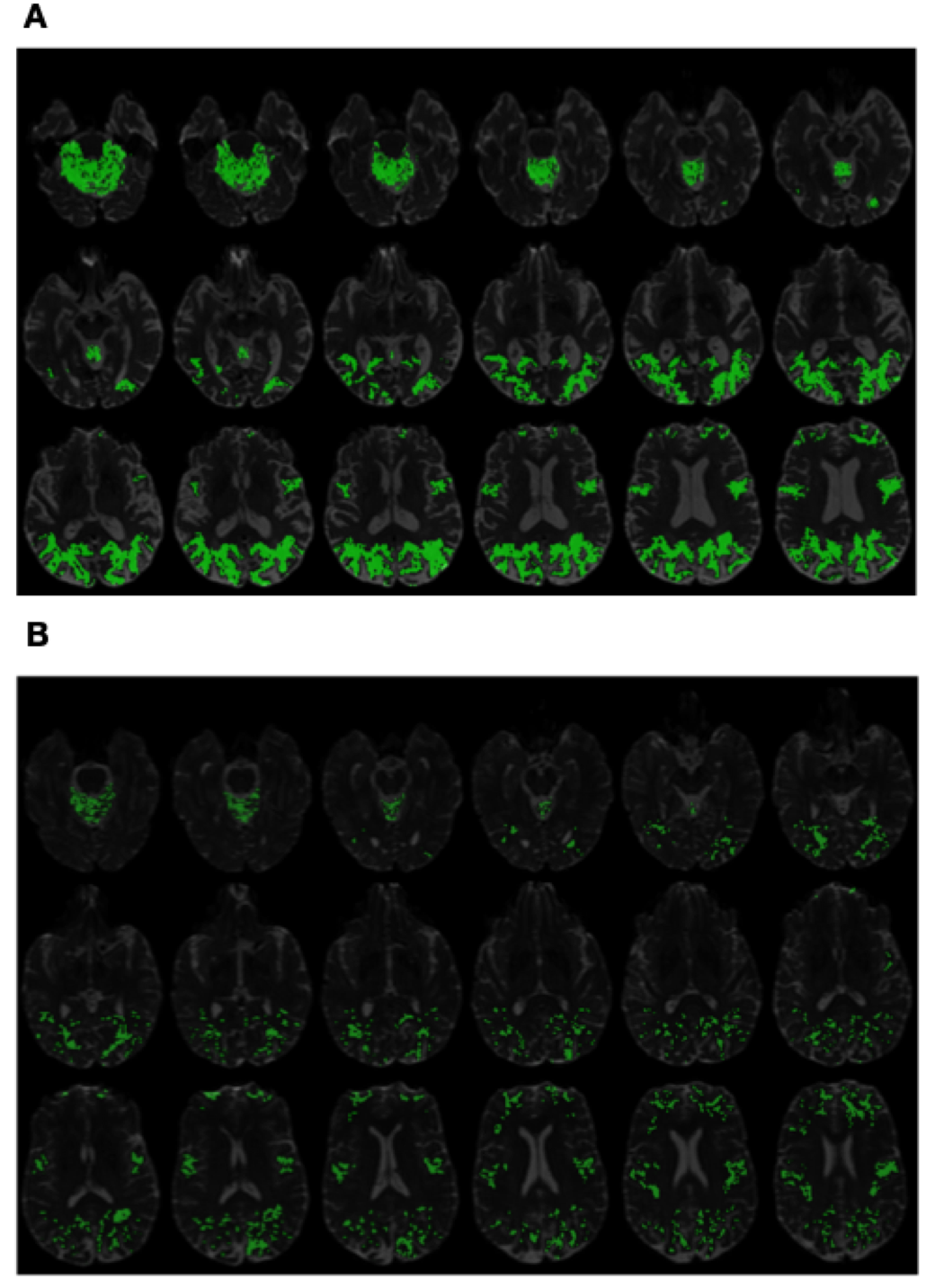}
\caption{(Subplot A) PSIC values of an AD subject superimposed on structural MRI scans. (Subplot B) PSIC values of a CN subject superimposed on structural MRI scans.}
\label{FGR4}
\end{center}
\end{figure}

\begin{figure}[!t]
\begin{center}
\includegraphics[width=5in]{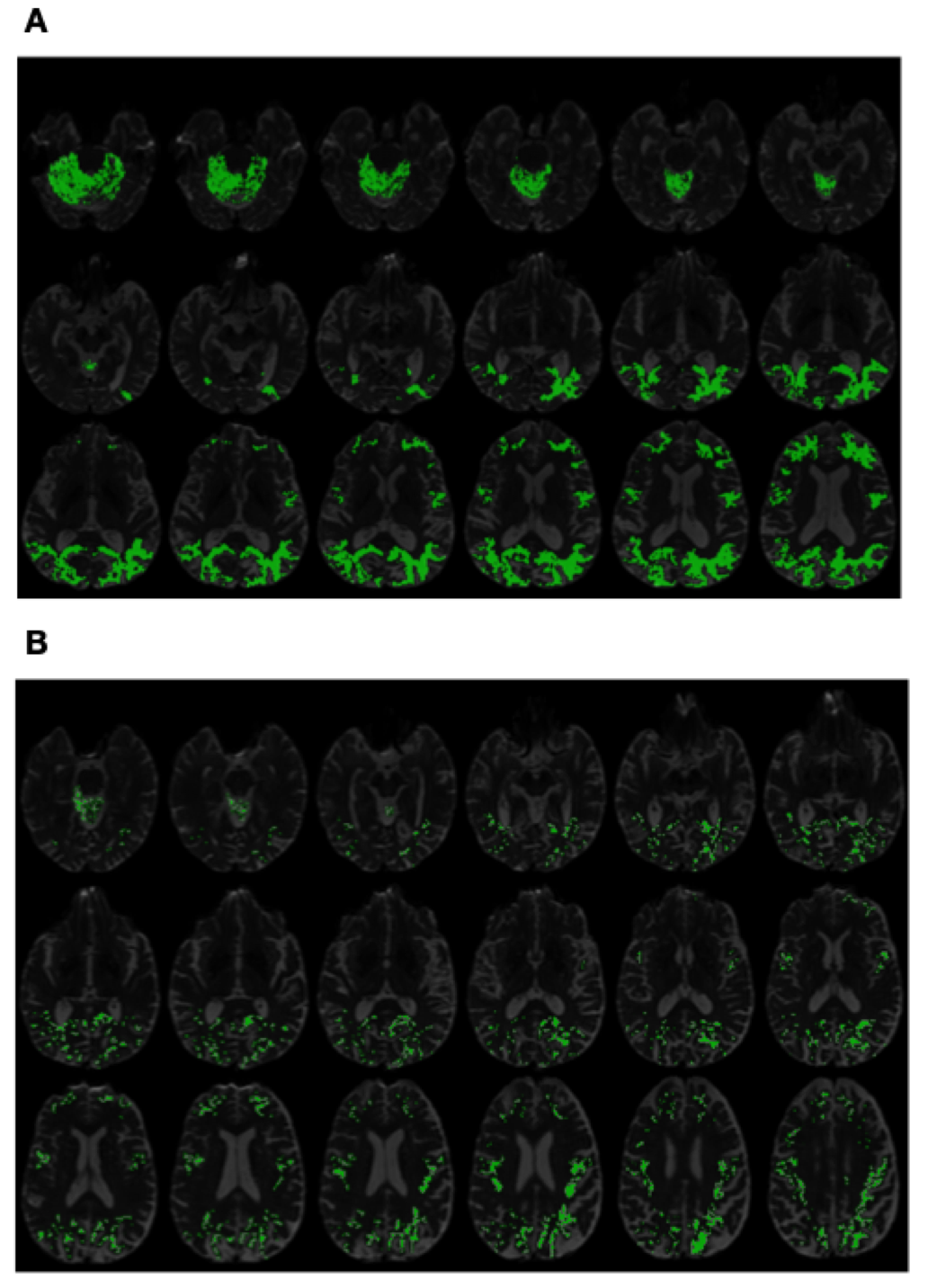}
\caption{(Subplot A) PSIC values of an AD subject superimposed on structural MRI scans. (Subplot B) PSIC values of a CN subject superimposed on structural MRI scans.}
\label{FGR5}
\end{center}
\end{figure}

Finally, Fig.~\ref{FGR4} and Fig.~\ref{FGR5} exemplify the use of PSIC in its capacity as imaging contrast. It should be emphasized that, as opposed to diffusion metrics, the values of PSIC have no physiological interpretation. Instead, PSIC could be viewed as a {\it pathology-specific} risk indicator, whose higher values reflect a higher probability of the brain to be affected by the disease. This property of PSIC is evident in Subplots A of Fig.~\ref{FGR4} and Fig.~\ref{FGR5} which show the PSIC-enhanced structural scans of two AD subjects. In this case, the spatial distribution of PSIC values appears to be both intense and spatially pervasive. On the other hand, Subplots B of the same figures show results for two CN subjects. One can see that, in this case, the magnitude and spatial spread of PSIC appear to be much more ``diluted".

Due to the preliminary nature of the present paper, an in-detail exploration of the spatial characteristics of PSIC as well as its correlation with underlying brain anatomy and its possible etiological explanations are left beyond the scope of this report. However, in view of the empirical evidence provided by Fig.~\ref{FGR4} and Fig.~\ref{FGR5}, it is reasonable to expect the proposed contrast mechanism to be ``worth a thousand words", both as an adjunct to establishing a confident diagnosis and as a means to facilitate {\it post hoc} discoveries.

% ---------------------------------------------------------------- SECTION 7 ------------------------------------------------------------------
\section{Discussion and Conclusions}\label{Sec7}
The main objective of this work has been to explore the potential of dMRI in application to early diagnosis of AD. As opposed to alternative means of medical imaging, the physics of dMRI happens to be uniquely suited for the detection and assessment of microscopic damage to brain tissue, which is known to precede the ensuing morphological changes in cortical grey matter due to AD \cite{Braak:1991aa, Kumar:2015aa}. This is what endows dMRI with the unique ability to detect the presence of neurodegeneration at its earliest pathological stages.

The proposed method has been derived based on the concept of {\it data-driven inference}, which allows overcoming some critical limitations of model-based analysis of diffusion signals, especially in situations with relatively small DTI datasets (i.e., when $K \lesssim 40$). The DNN-based classifier designed this way has been rendered independent of any mechanistic assumptions (and, thus, of their limitations). Instead, it has been optimized based on known diagnostic outcomes, resulting in the phenomenological mechanism that establishes a {\it direct} correspondence between DTI data and a quantitative measure of health risks due to AD.

The proposed DNN has been designed to process spatially localized segments of DTI data, i.e., 4-D ``diffusion cubes" corresponding to different spatial coordinates. The local definition of input DC samples has served two important purposes. First, it gave the means to use the output scores as a spatially dependent  ``risk indicator", which has been referred to as PSIC (in view of its purposive specificity to suspected pathology). Second, the same locality made it possible to collect tens of thousands of training samples from as few as only 40 DTI datasets. More importantly, the training data thus obtained have been sufficient for reliable optimization of the network parameters. Needless to say, such a result would have been impossible to attain, had, in accordance with established practice, each of the datasets been dealt with as a {\it single} observation. Thus, the proposed methodology can be particularly advantageous in situations when larger sets of training data are not available.

It goes without saying; the present work has barely scratched the surface of the possibilities offered through the combination of dMRI measurements and DL, with many important questions left yet to be addressed. In particular, although sufficient as a ``proof-of-concept" in comparative studies, the problem of DTI-based stratification of CN and AD subjects is clearly of limited practical importance. Thus, to further attest to its viability, the proposed method needs to be extended to the classification of multiple diagnostic groups. Along this direction of research, a particularly enticing question would be to find correspondence between PSIC and different pathological stages of AD \cite{Braak:1991aa}. In a similar vein, it would also be interesting to apply the proposed solution to the problem of classification of various subtypes of MCI \cite{Petersen:2016aa}.

Addressing more complex classification problems is likely to require a proportional increase in the complexity of the DNN. Thus, in the case of simple CN/AD stratification, it was neither necessary to involve more complex data nor to extend the network architecture beyond a basic feed-forward configuration. Moreover, the minimality of the employed DNN architecture (with the total number of trainable parameters equal to 50,376) has been instrumental in preventing overfitting during its training on the small set of 40 subjects. In more complicated clinical scenarios, however, the architecture could be extended through, e.g., processing the spatial neighbourhoods $\Omega_L$ of different sizes, with a corresponding increase in the number of hidden layers. Another way to enhance the predictive power of the DNN would be to take advantage of dMRI data acquired at {\it multiple} $b$-values (as has been done in the ADNI-III study). Needless to add, all such extensions would come with an increase in the number of network parameters, which should be accompanied by a {\it pro-rata} increase in the size of training data.

Finally, although working with SH coefficients is not the only way to ``planarize" the operation of spherical convolution, it offers an important advantage in the context of between-site variability of classification scores. Specifically, due to discrepancies in the design and settings of MRI scanners, similar dMRI signals acquired at different sites are not uncommon to have notably different spectral characteristics (which is often the reason behind conflicting reports in clinical applications of dMRI). This problem has been addressed by a range of approaches, among which a particularly effective way to counteract the effects of between-site variability is lent by means of {\it spectral data harmonization} of the SH coefficients of DTI data, as detailed in \cite{Mirzaalian:2018aa}. This approach suggests a straightforward way of combining the proposed method with the compatible means of normalization, which should render its performance consistent across different clinical sites. This expectation, however, still needs to be validated via proper experimental studies, which constitutes another objective of our future research.

\bibliographystyle{unsrt}
\bibliography{deep.bib}

\end{document}